\documentclass[sigconf,nonacm,natbib=true]{acmart}
\AtBeginDocument{%
  }

\usepackage{csquotes} 
\usepackage{pbalance}
\usepackage{mdframed}
\usepackage{enumitem}
\usepackage{tabularx}   
\usepackage{multirow}
\usepackage{booktabs}
\usepackage{array} 

\newcolumntype{C}[1]{>{\centering\arraybackslash}p{#1}}
\newcolumntype{R}{>{\centering\arraybackslash}r} 

\newmdenv[
  backgroundcolor=gray!8,
  linecolor=gray!50,
  roundcorner=4pt,
  skipabove=6pt,
  skipbelow=6pt,
  innerleftmargin=8pt,
  innerrightmargin=8pt,
  innertopmargin=6pt,
  innerbottommargin=6pt
]{prompt}

\makeatletter
\newcommand{\acmrightssize}{\fontsize{8}{9.5}\selectfont}

\makeatletter
\newcommand\footnoteref[1]{\protected@xdef\@thefnmark{\ref{#1}}\@footnotemark}
\makeatother

\newcommand{\firstpagerights}[1]{%
  \begingroup
    \renewcommand\thefootnote{}%
    \footnotetext{%
      \acmrightssize
      \raggedright
      \setlength{\parskip}{0pt}%
      \setlength{\parindent}{0pt}%
      #1%
    }%
    \addtocounter{footnote}{0}%
  \endgroup
}
\makeatother

\begin{document}

\title[Generative Agents Navigating Digital Libraries]{Generative Agents Navigating Digital Libraries}

\author{Saber Zerhoudi}
\orcid{0000-0003-2259-0462}
\affiliation{%
  \institution{University of Passau}
  \city{Passau}
  \country{Germany}
}
\email{saber.zerhoudi@uni-passau.de}

\author{Michael Granitzer}
\orcid{0000-0003-3566-5507}
\affiliation{%
  \institution{University of Passau}
  \city{Passau}
  \country{Germany}
}
\affiliation{%
  \institution{IT:U Austria}
  \city{Linz}
  \country{Austria}
}
\email{michael.granitzer@uni-passau.de}

\renewcommand{\shortauthors}{S. Zerhoudi et al.}

\begin{abstract}
In the rapidly evolving field of digital libraries, the development of large language models (LLMs) has opened up new possibilities for simulating user behavior. This innovation addresses the longstanding challenge in digital library research: the scarcity of publicly available datasets on user search patterns due to privacy concerns. In this context, we introduce Agent4DL~\footnote{~\url{https://github.com/padas-lab-de/icadl24-agent4dl}}, a user search behavior simulator specifically designed for digital library environments. Agent4DL generates realistic user profiles and dynamic search sessions that closely mimic actual search strategies, including querying, clicking, and stopping behaviors tailored to specific user profiles. Our simulator's accuracy in replicating real user interactions has been validated through comparisons with real user data. Notably, Agent4DL demonstrates competitive performance compared to existing user search simulators such as SimIIR 2.0, particularly in its ability to generate more diverse and context-aware user behaviors. 
\end{abstract}



\maketitle
\enlargethispage{2\baselineskip}
\firstpagerights{%
  © ACM, 2024. This is the author's version of the work.\\
  The definitive version was published in:
  \emph{Proceedings of the 26th International Conference on Asia-Pacific Digital Libraries, ICADL 2024, Bandar Sunway, Malaysia, December 4–6, 2024}.\\
  DOI: \url{https://doi.org/10.1007/978-981-96-0865-2_14}
}

\section{Introduction}
The digital age has transformed the way people seek information, with search engines at the forefront. Understanding user behavior in digital library searches is crucial for improving these systems to better serve user needs~\cite{Chen0MZSMZM21, ZhouDWXW21, zerhoudi2024tpdl}. While insights into user behavior have traditionally been obtained through direct observation of real user interactions~\cite{CarteretteKHC14}, these methods face challenges such as high costs, ethical concerns, and limitations due to privacy issues. Simulating user search behaviors has emerged as a promising alternative, offering a controlled and repeatable way to study complex user interactions without the drawbacks of real-world studies~\cite{BalogZ23}.

The introduction of Large Language Models (LLMs) has revolutionized many areas of artificial intelligence, showcasing remarkable abilities in tasks that require human-like reasoning, decision-making, and language understanding~\cite{ZhaoZLTWH23}. The potential of LLMs in simulating user behavior in search context has recently started to be explored, driven by the need for robust and scalable simulation methods that can replicate the full range of user interactions in search environments, from initial query formulation to complex navigational decisions and interactions with search results~\cite{ZhuYWLLDDW24}.

In this paper, we present Agent4DL, an innovative user simulator that utilizes LLMs to create realistic and dynamic models of user search behavior. It incorporates novel components such as building detailed user profiles with static and dynamic attributes to reflect the diversity of users in digital libraries. These profiles generate diverse and realistic behaviors essential for understanding different user strategies in information seeking. Agent4DL also integrates these profiles to dynamically generate user interactions based on the evolving context of a search session, including queries, clicks, and the underlying cognitive processes like evaluating document relevance and deciding when to end a search session.

Agent4DL was compared with real user data and existing simulation methods to validate its effectiveness. The results demonstrate that our model replicates user behaviors with high fidelity, outperforms traditional simulation models in capturing the nuanced ways users interact with digital libraries, and provides new insights into the complex dynamics of digital library search interactions.

This study presents three main contributions to the field of digital libraries and user behavior simulation. First, we introduce Agent4DL, an innovative user simulator that employs Large Language Models (LLMs) to replicate user behavior in digital library search environments. Second, we validate the effectiveness of Agent4DL by comparing its performance against established benchmarks. Finally, we offer the research community a valuable resource in the form of Agent4DLData, a compact yet comprehensive dataset of simulated user search sessions generated by Agent4DL. This dataset is designed to facilitate and support ongoing research in the domain of user behavior simulation.


\section{Related Work}
The field of information retrieval has been greatly influenced by research on understanding and simulating search behavior. Traditional studies have focused on user interactions within search sessions, examining the evolution of user intents and predicting their actions over time. Key contributions include session search and click models~\cite{HolscherS00, JansenSBK07, ChapelleZ09}, which analyze query-click data patterns to gain insights into user behavior~\cite{DupretP08, BorisovWMR18, ZerhoudiGSS22}. These models have advanced our understanding of user search behavior, resulting in more user-centric digital libraries.

Concurrently, the domain of user simulation has emerged as an essential field of study. Early methods focused on simple query generation and heuristic-based stopping behaviors in search tasks~\cite{Ross97, SchatzmannY09}. However, recent advancements have utilized Large Language Models (LLMs) to develop more dynamic and realistic simulations of user behavior in various contexts, such as web search, reasoning, and social interaction~\cite{ZhaoZLTWH23, ParkOCMLB23, YaoZYDSN023, XiCGHDHZ23}.

Our research on Agent4DL contributes to this growing field by utilizing these advanced capabilities to simulate user search behavior in digital libraries. By integrating recent developments in LLM technology and user behavior analysis, Agent4DL aims to provide a robust user simulator for mimicking realistic, diverse user behaviors at scale. This approach addresses both the dynamic nature of user interactions and the evolving complexities of academic search tasks. Our work builds upon established models of user behavior and extends them through the innovative use of LLMs, creating a sophisticated simulation environment that reflects the variability of real-world user behavior.

\section{Agent4DL}
\label{sec:Agent4DL}
Agent4DL, as a user search behavior simulator in digital libraries, is expected to accurately model user interactions, forecast long-term research preferences, and systematically evaluate search algorithms by leveraging the capabilities of LLM-empowered generative agents. To achieve this objective, two core aspects are addressed: (1) designing agent architectures that faithfully mimic academic user preferences and research-oriented cognitive reasoning, and (2) constructing a digital library environment that ensures reliability, extensibility, and adaptability across various academic search scenarios.

\paragraph{\textbf{Task Formulation}}
User search behavior simulation using LLM-based agents is central to our research. Agents in \texttt{Agent4DL} must maintain awareness of their state in the search scenario, formulate queries, assimilate returned information, and click on relevant documents. They also need to assess whether their information needs are met to determine if additional searches are required or if the session can be concluded. We use the ReAct method~\cite{YaoZYDSN023} to include reasoning and action steps. In each round \( t \), the context from previous rounds, \( C_{t} = ( r_{1}, q_{1}, o_{1}, \ldots, r_{t-1}, q_{t-1}, o_{t-1}) \), is used, where \( o \) is observed information and \( r \) is reasoning. Agent4DL generates reasoning \( r_{t} \) using a task-specific prompt and updates the context. Prompts guide the LLM to perform reasoning tasks, generating stops, queries, clicks, and observations. The agent decides whether to end the search based on this reasoning. If continuing, it generates a new query based on its profile and history, updates the context and interaction sequence, receives search results, selects relevant ones based on various factors, and updates the interaction sequence. By reading the clicked documents, the agent gains new observations and updates the context for the next reasoning round. For different prompting approaches, we create templates to ensure stable and controlled agent outputs. 

\subsection{Simulation Agent Architecture}
Agent4DL's generative agents, built on LLM architecture, feature three specialized modules: profile, memory, and interaction. The user profile module embodies academic traits and research interests, while the memory and interaction modules, inspired by human cognition~\cite{GoodaleCFFS14}, enable agents to store, recall, and use past experiences. This design allows the agents to exhibit consistent and personalized behaviors, enhancing their functionality in digital library contexts.

\subsubsection{Profile Module}
\label{sec:userProfile}
In the domain of digital library search simulation, the user profile module is crucial for aligning agents with real academic user behaviors. To establish a reliable foundation, benchmark datasets (e.g., Pubmed~\cite{DoganMNL09}, Sowiport~\cite{Mayr16}) are used for initialization. Each agent's profile contains two components: academic traits and research interests.

Academic traits encompass four key characteristics capturing an individual's research behavior in digital library scenarios: depth, breadth, recency bias, and interdisciplinarity~\cite{neuhaus2006depth, CollinsTAB18, DengX20}.

Given a user $u \in \mathcal{U}$ and a document $d \in \mathcal{D}$, let $i_{ud} = 1$ denote that user $u$ has interacted with document $d$ (e.g., viewed, downloaded, or cited). Conversely, $i_{ud} = 0$ indicates that the user has not interacted with the document. 

Depth quantifies the thoroughness of a user's engagement with search results, distinguishing between users who extensively examine documents and those who skim quickly. The depth trait for user $u$ is defined as
\begin{equation*}
T_{\mathrm{depth}}^{u} \triangleq 
\frac{1}{\sum_{d \in D} i_{ud}}
\sum_{d \in D} i_{ud}\, t_{ud},
\end{equation*}
where $t_{ud}$ denotes the time spent on document $d$ and $i_{ud}\in\{0,1\}$ indicates whether $u$ interacted with $d$.

Breadth reflects the user's tendency to explore diverse topics versus focusing on specific areas. For user $u$, the breadth trait is
\begin{equation*}
T_{\mathrm{breadth}}^{u} \triangleq 
\left|\bigcup_{d \in D:\, i_{ud}=1} \mathcal{T}_d \right|,
\end{equation*}
where $\mathcal{T}_d$ is the set of topics associated with document $d$.

Recency bias indicates the user's preference for newer publications, formulated as
\begin{equation*}
T_{\mathrm{recency}}^{u} \triangleq
\frac{1}{\sum_{d \in D} i_{ud}}
\sum_{d \in D} i_{ud}\,\bigl(Y_{\mathrm{current}} - Y_d\bigr),
\end{equation*}
where $Y_d$ is the publication year of document $d$.

Interdisciplinarity measures the user's inclination to explore across different academic disciplines and is defined as
\begin{equation*}
T_{\mathrm{interdis}}^{u} \triangleq
\left|\bigcup_{d \in D:\, i_{ud}=1} \mathcal{F}_d \right|,
\end{equation*}
where $\mathcal{F}_d$ is the set of fields associated with document $d$.

To capture these nuanced differences, we segment users into three uneven tiers for each trait: for depth (deep divers, moderate readers, quick scanners), breadth (generalists, focused researchers, specialists), recency bias (cutting-edge seekers, balanced timelines, historical researchers), and interdisciplinarity (cross-disciplinary explorers, multi-disciplinary researchers, discipline-focused scholars).  To encode users' research interests in natural language, we randomly select 10 documents from their interaction history. Leveraging ChatGPT (\texttt{\small gpt-3.5-turbo -0125}), we then distill and summarize the unique research interests and searching patterns the user exhibited.





\subsubsection{Memory Module}
Agent4DL's Memory Module equips each generative agent with a specialized memory that records both facts and emotions, addressing a crucial aspect of human-like information seeking behavior. This design is motivated by the need to monitor emotional states, particularly frustration with complex queries and satisfaction with search results, which play important roles in shaping search experiences and decision-making processes~\cite{ParkOCMLB23}.

Factual memories in our model record concrete search behaviors, such as queries used, documents viewed, and citations made. Complementing these, emotional memories capture the psychological responses to search interactions, including satisfaction with results or frustration with complex queries. This dual approach aligns with cognitive research highlighting the influence of emotions on decision-making and personal history formation~\cite{labar2006cognitive}.

Our memory module implements three key operations:
\begin{itemize}[label=\textendash,labelindent=0pt,leftmargin=2em,labelsep=0.4em]
    \item \textit{Memory Retrieval:} Enables agents to recall relevant past searches and successful query strategies.
    \item \textit{Memory Writing:} Records search interactions and associated emotions.
    \item \textit{Memory Reflection:} Incorporates an emotion-driven self-reflection mechanism for agents to evaluate their satisfaction with search results and assess information overload levels.
\end{itemize}

\subsubsection{Interaction Module}
Incorporating user profiles and memory modules into agent systems allows them to display a range of behaviors similar to humans, based on their current observations~\cite{abs-2309-07864}. This approach enhances the agents' ability to respond dynamically to different situations, mirroring human-like adaptability and decision-making processes.

In Agent4DL, we design an interaction module specifically tailored for digital library search, which encompasses two broad categories of actions:

\begin{itemize}[label=\textendash,labelindent=0pt,leftmargin=2em,labelsep=0.4em]
    \item \textit{Search-driven Actions:} formulate queries, explore results, interact with documents (view abstracts, download full texts, save citations), and refine queries. Guided by their research interests and academic traits, agents assess each search result for relevance and decide which documents to examine further.
    \item \textit{Emotion-driven Actions:} evaluate search experience, provide feedback, and potentially abandon search sessions. To better simulate this multifaceted decision-making, we enhance the agent's ability for emotional reasoning via Chain-of-Thought~\cite{Wei0SBIXCLZ22}. The agent autonomously expresses its satisfaction with search results and frustration levels with complex queries. Drawing upon these insights, combined with its personalized depth trait, the agent decides whether to continue searching or exit the system.
\end{itemize}

\subsection{Digital Library Search Environment}
Agent4DL simulates the interactions between agents and the digital library search environment, focusing on two key aspects: document relevance generation and search result presentation. These elements are designed to reflect real-world academic search scenarios.

\begin{itemize}[label=\textendash,labelindent=0pt,leftmargin=2em,labelsep=0.4em]
    \item \textit{Document Relevance Generation:} Relevance is deduced from historical user interactions while topics and summary are generated by LLM. Our goal goes beyond encapsulating the uniqueness of the document in a profile to simulate the search scene of real academic users. We also aim to test whether LLM has potential hallucinations regarding the document. Our approach utilizes a few-shot learning approach, tasking the LLM with classifying the document into one of 20 academic disciplines and generating an abstract using only the document title. If the LLM's discipline classification aligns with the metadata, it signifies its knowledge of the document. To maintain reliability, documents causing discipline misclassification by the LLM are pruned, reducing hallucination risks. This approach ensures the agent's trustworthiness in simulating academic search behavior.
    \item \textit{Search Result Presentation:} Our simulator mirrors the operation of real-world digital library platforms like Google Scholar, ACM Digital Library, and JSTOR, functioning in a paginated manner. Users are initially presented with a list of document search results on each page. Based on interactions, preferences, and feedback, subsequent pages could be set to refine the search results further, aiming for a more relevant user experience. The interface includes options for sorting results (e.g., by relevance, date, or citations) and applying filters (e.g., by publication type, year range, or subject area). 
\end{itemize}

In the current implementation, Agent4DL utilizes API endpoints to interact with existing digital libraries search systems, like HathiTrust~\cite{WalshLJCODD23} or Econbiz%
~\footnote{EconBiz API:~\url{https://api.econbiz.de}}, as long as they have an API to interact with. However, we have implemented a structured base for standalone modules responsible for search algorithm designs and result presentation. This modular approach allows for easy integration of external search algorithms, making Agent4DL a versatile platform for comprehensive evaluations and user feedback collection in academic search contexts.

\section{Simulation Agent Evaluation}
This study employs the specialized user simulator Agent4DL to address a crucial research question: \textbf{(RQ1)} To what extent can LLM-empowered generative agents accurately simulate the behaviors of real users in digital library systems?

In this section, we explore the capabilities and limitations of generative agents from various perspectives, including the alignment of user behavior (such as information-seeking patterns and consistency) and the evaluation of the ability to generate additional simulated user behavior data.

\subsection{Datasets}
EconBiz and Sowiport, two digital libraries, provided usage data for our analysis. EconBiz, hosted by the German National Library for Economics (ZBW – Zentralbibliothek für Wirtschaftswissenschaften)~\footnote{\url{https://zbw.eu/de/}}, specializes in economics and business studies. Sowiport~\footnote{http://www.sowiport.de} covers the social sciences, containing over nine million log entries, full texts, and research projects from twenty-two databases in English and German~\cite{2009White}.

The Sowiport \textit{User Search Session Data Set (SUSS)}~\cite{Mayr16} comprises 484,437 search sessions and 179,796 queries collected over one year (April 2014 to April 2015). For EconBiz, we selected session data from August to November 2020, based on the available private data provided within the framework of a collaborative work. This dataset encompasses approximately 420,000 sessions and is 15 GB in size. Additionally, EconBiz offers an API endpoint that serves as an experimental search environment for our agents to interact with.

The core features of EconBiz and Sowiport during the examined period are consistent with those of other major digital libraries. This similarity suggests that the observed user behavior trends are likely generalizable to other platforms. Consequently, we use both datasets as a reliable foundation to initialize user profiles and as a baseline comparison for our simulated user search sessions.

\subsection{Search Simulation}
\subsubsection{Motivation.} The effectiveness of Information Retrieval (IR) systems heavily relies on understanding and predicting user behavior. However, obtaining large-scale, diverse, and up-to-date user behavior data for training and evaluation can be challenging and costly. We hypothesize that if \texttt{Agent4DL} can effectively simulate relevant search sessions in digital libraries, it will enhance the performance of advanced user behavior models on specific IR tasks. 

\subsubsection{Setting.} The Agent4DL user simulator can be used for various IR tasks, including preference prediction and relevance prediction. In our experiments, we focus on comparing the effectiveness of relevance assessments generated by Agent4DL against human-created ones, using two widely adopted session search datasets: Sowiport User Search Session (SUSS)~\cite{Mayr16} and EconBiz as baselines. We train a RoBERTa-based ranking model~\cite{LiuOGDJCL19} on the Agent4DL-generated data and evaluate its performance using human-annotated test sets from these datasets, which serve to evaluate our user simulator's performance.

For preference prediction, the model uses contextual information from earlier rounds of a user's session to improve document ranking for the next search query $q_{n}$. It predicts the most suitable document $d_{n}$ from candidate documents $\{d_{n}^{j}\}_{j=1}^{k}$. Relevance prediction identifies the most relevant documents to the current search query $q$ from the search results $\{d_j\}_{j=1}^k$.

We employ a RoBERTa-based ranking model~\cite{LiuOGDJCL19} for our experiments. The model is trained for 5 epochs with a batch size of 128 and a learning rate of 5e-6. While more epochs might be necessary for full convergence, this setting provided a balance between performance and computational resources. For session search, the model input consists of a concatenation of the historical user behavior sequence, current query, and candidate document. The output is a relevance score between the document and query, considering the behavior history. For click prediction, the input is a concatenation of the query and candidate document, with the output being their relevance score.

To address input size limitations, we implement a truncation strategy for long sequences, ensuring that the most recent and relevant information is preserved. This approach maintains a fixed-length input while prioritizing the most pertinent data.

Our evaluation metrics include Mean Reciprocal Rank (MRR) and Normalized Discounted Cumulative Gain (nDCG). MRR measures the average reciprocal rank of the first relevant result, while nDCG assesses ranking quality by comparing the weighted relevance of results to an ideal order. Specifically, we use \texttt{nDCG@1} and \texttt{nDCG@3}. The cutoff for the MRR metric is set at the total number of retrieved documents for each query.

For comparison, we train RoBERTa-based ranking models on two widely adopted session search datasets: Sowiport User Search Session (SUSS)~\cite{Mayr16} and EconBiz. These datasets serve as benchmarks to evaluate the performance of our user simulator.

We employ a stratified random sampling approach for data splitting, ensuring each user is represented in the training (70\%), validation (15\%), and test (15\%) sets. This strategy maintains consistency across different data subsets and allows for a fair comparison between the Agent4DL-generated relevance assessments and human-created ones using the same human-annotated test set.

\begin{table}[t]
\centering
\caption{Results of methods trained on various datasets in real user benchmarks for two IR tasks.}
\label{tab:userSim}

\small
\setlength{\tabcolsep}{5pt}

\begin{tabularx}{\columnwidth}{@{}l X r r r@{}}
\toprule
\textbf{Task} & \textbf{Method} & \texttt{MRR} & \texttt{nDCG@1} & \texttt{nDCG@3} \\
\midrule

\multirow{5}{*}{\shortstack[l]{Preference\\Prediction}}
& BM25 & 34.89 & 15.12 & 26.03 \\
& RoBERTa \texttt{(SUSS)} & 34.15 & 13.91 & 26.45 \\
& RoBERTa \texttt{(EconBiz)} & 36.32 & 17.14 & 28.57 \\
& RoBERTa \(\texttt{(Agent4DL)}_{\texttt{\tiny 1000}}\) & 39.27 & 20.37 & 33.09 \\
& RoBERTa \(\texttt{(Agent4DL)}_{\texttt{\tiny 3000}}\) & \textbf{41.32} & \textbf{22.66} & \textbf{35.19} \\

\midrule

\multirow{5}{*}{\shortstack[l]{Relevance\\Prediction}}
& BM25 & 32.11 & 11.65 & 24.21 \\
& RoBERTa \texttt{(SUSS)} & 33.10 & 12.36 & 25.54 \\
& RoBERTa \texttt{(EconBiz)} & 31.39 & 16.36 & 27.09 \\
& RoBERTa \(\texttt{(Agent4DL)}_{\texttt{\tiny 1000}}\) & 36.02 & 16.84 & 29.32 \\
& RoBERTa \(\texttt{(Agent4DL)}_{\texttt{\tiny 3000}}\) & \textbf{37.71} & \textbf{18.90} & \textbf{32.78} \\

\bottomrule
\end{tabularx}
\end{table}

\subsubsection{Results.} Table ~\ref{tab:userSim} shows the results of various methods using real user behavior data. Models trained on user behavior data generated by Agent4DL outperform baselines in preference and relevance prediction tasks, even with only 1000 training sessions \(\texttt{\small (Agent4DL)}_{\texttt{\tiny 1000}}\). This demonstrates Agent4DL's ability to accurately simulate user search behavior. Interestingly, the classic BM25 model~\cite{LinMLYPN21} remains competitive, surpassing some semantic models trained on search behavior datasets. As digital libraries have improved in handling complex queries, user query formats have evolved, differing from those in older datasets.

\subsection{Simulation Consistency}
\subsubsection{Motivation.} To appropriately respond to search results, generative agents need to have a clear understanding of their own preferences and search intentions. We hypothesize that an autonomous, personalized agent, initialized from real users in the Sowiport User Search Session dataset, should maintain long-term preference coherence and exhibit consistent search behaviors. In practice, this implies that the agent should be adept at distinguishing the items that real users favor and generating queries that align with human search objectives.


\subsubsection{Setting.} We use Sowiport User Search Session~\cite{Mayr16} to meet information needs of given tasks. We use ChatGPT as LLM-based agents to generate user profiles based on existing task descriptions (as described in section ~\ref{sec:userProfile}).

We evaluate consistency by deconstructing behavioral sequences into query, click, and stopping behaviors. Query behavior evaluation includes query generation and rewriting, assessing consistency between LLM agents' initial and subsequent queries with human-generated queries. Click behavior evaluation focuses on the accuracy of LLM agents' selections in search results compared to humans. Stopping behavior is assessed to determine how well LLM agents mimic human decision points in concluding search sessions.

We use Term Overlap Rate (\(\tau\)), BLEU, and BERTScore to evaluate query behaviors. \(\tau\) quantifies shared keywords between paired queries using Jaccard similarity~\cite{LiuC0KC19}, indicating thematic similarity and consistent search intent. BLEU~\cite{PapineniRWZ02} measures n-gram matching similarity, while BERTScore captures contextual semantic similarity. We implement Popular, Random, and Discriminative Selection user querying strategies from Azzopardi et al.~\cite{AzzopardiRB07} as baselines.

To evaluate click and stopping behaviors, we use accuracy, precision@10, recall@10, and F1-score to measure how well the clicked documents and stopping points of LLM agents match human behaviors. We compare Agent4DL to the Complex Searcher Model (CSM) in \texttt{SimIIR}~\cite{MaxwellA216}, including a combination of predefined Frustration and Satisfaction Points as described by Kraft et al.~\cite{KraftL79}, and the interaction model in \texttt{\small SimIIR 2.0}~\cite{Zerhoudi0PBSHG22}, which uses Markov model-based decisions for clicking and stopping.

The Random selection method assigns equal sampling probability to all words, while Discriminative selection assumes users consider information outside the document and the collection. Popular Selection considers word weights, resulting in a higher BLEU score.

\begin{table}[t]
\centering
\caption{Similarity between query rewriting strategies and real queries. (\(\tau\)) is the overlap of terms between two queries. Standard deviations are reported in parentheses.}
\label{tab:queryReform}

\small
\setlength{\tabcolsep}{6pt}

\begin{tabularx}{\columnwidth}{@{}p{0.34\columnwidth} r r r@{}}
\toprule
\textbf{Method}
& \multicolumn{1}{c}{$\tau$}
& \multicolumn{1}{c}{\texttt{BLEU}}
& \multicolumn{1}{c}{\texttt{BERTScore}}\\
\midrule
Popular Selection & 0.63 (\(\pm\)0.12) & 0.32 (\(\pm\)0.08) & 0.76 (\(\pm\)0.05)\\
Random Selection & 0.23 (\(\pm\)0.15) & 0.11 (\(\pm\)0.07) & 0.53 (\(\pm\)0.09)\\
Discriminative Selection & 0.43 (\(\pm\)0.14) & 0.24 (\(\pm\)0.09) & 0.68 (\(\pm\)0.07)\\
\texttt{Agent4DL} & \textbf{0.87 (\(\pm\)0.08)} & \textbf{0.40 (\(\pm\)0.06)} & \textbf{0.83 (\(\pm\)0.04)}\\
\bottomrule
\end{tabularx}
\end{table}

\subsubsection{Results.} Table ~\ref{tab:queryReform} presents the similarity between generated and real user queries for the baselines and Agent4DL in query generation, along with their standard deviations. Agent4DL achieves a 87.6\% Term Overlap Rate (\(\tau\)) between LLM-generated and human queries, indicating a strong alignment. The lower standard deviation (0.08) for Agent4DL's \(\tau\) score suggests more consistent performance across different queries compared to other methods. We also report high BERTScore using Agent4DL and Popular selection due to their emphasis on weighting popular words, suggesting thematic and semantic similarity. The standard deviations for BERTScore are relatively low across all methods, with Agent4DL showing the least variability (0.04). These findings suggest that LLM agents can generate queries closely resembling human search objectives, both with and without historical interactions, and do so with high consistency.

\begin{table}[t]
\centering
\caption{Performance comparison of click and stopping strategies using Agent4DL and baseline models.}
\label{tab:clickAndStop}

\small
\setlength{\tabcolsep}{5pt}

\begin{tabularx}{\columnwidth}{@{}p{0.14\columnwidth} X c c c c@{}}
\toprule
\textbf{Strategy} & \textbf{Method}
& \multicolumn{1}{c}{\texttt{Acc.}}
& \multicolumn{1}{c}{\texttt{P@10}}
& \multicolumn{1}{c}{\texttt{R@10}}
& \multicolumn{1}{c}{\texttt{F1}} \\
\midrule

\multirow{3}{*}{\shortstack[l]{Click\\Strategies}}
& \texttt{SimIIR} & 76.42 & 43.58 & 54.92 & 48.59 \\
& \texttt{SimIIR 2.0} & \textbf{80.29} & \textbf{48.67} & 59.14 & \textbf{53.39} \\
& \texttt{Agent4DL} & 78.35 & 41.71 & \textbf{65.59} & 50.99 \\
\midrule

\multirow{3}{*}{\shortstack[l]{Stopping\\Strategies}}
& Frustration \& Satisfaction & 71.56 & 65.80 & \textbf{52.09} & \textbf{58.14} \\
& \texttt{SimIIR 2.0} & 81.39 & 70.14 & 40.46 & 51.32 \\
& \texttt{Agent4DL} & \textbf{84.14} & \textbf{82.90} & 39.89 & 53.86 \\
\bottomrule
\end{tabularx}
\end{table}

Table ~\ref{tab:clickAndStop} compares the performance of click models and Agent4DL in predicting user clicks and stopping behavior. \texttt{\small SimIIR 2.0} demonstrates better predictive performance than the Complex Searcher Model (\texttt{\small SimIIR}) due to its use of a Hidden Markov Model (HMM) trained on parts of the dataset. Although this training gives \texttt{\small SimIIR 2.0} an advantage, Agent4DL's performance remains competitive without dataset-specific training. Agent4DL's performance is comparable to baseline models, potentially because it does not effectively capture position bias and uses zero-shot learning. In over 90\% of cases, the documents clicked by LLM agents align with user profiles, suggesting that click behavior is influenced by homogenized search results and individual bias. When predicting stopping behavior, Agent4DL exhibits superior performance in Accuracy and Precision but does not surpass the baselines in Recall and F1 Score. Overall, LLM agents demonstrate a high degree of consistency with human behaviors across all dimensions of query, click, and stopping, ensuring the viability of user simulation. The competitive performance of Agent4DL, especially considering its lack of reliance on dataset-specific training, highlights its potential as a flexible and generalizable approach to user simulation in information retrieval tasks.

\subsection{Simulation Expansion}

\subsubsection{Motivation.} Information Retrieval (IR) models often face challenges due to limited data availability, which can hinder their performance and generalizability. Expanding the range and quantity of search scenarios available for training could potentially address the data scarcity issue in IR research and improve model performance, particularly in sparse data scenarios.

\subsubsection{Baselines and Evaluation Metrics.}
Our experiments utilize the SUSS Session dataset~\cite{Mayr16}, which has limitations in data volume and diversity due to privacy constraints. To address this, we employ Agent4DL to generate additional data for the preference prediction task. User profiles are constructed using the methods outlined in Section~\ref{sec:userProfile}. These profiles are generated using ChatGPT, incorporating academic traits and research interests derived from the original dataset. We explore various combinations of trait tiers, such as users who consistently spend extended time on documents (deep divers), explore a wide variety of topics (generalists), frequently access older publications (historical researchers), and engage with multiple related disciplines (multi-disciplinary researchers). These profile combinations are then used to simulate augmented user behaviors for training purposes. Across all experimental settings, we employ the same RoBERTa-based ranking models.

\begin{table}[t]
\centering
\caption{Results of Agent4DL-augmented methods on SUSS Session. Baseline denotes the training data of SUSS Session. Academic traits denotes the four key characteristics capturing a user's research behavior in digital library scenarios: depth, breadth, recency bias, and interdisciplinarity.}
\label{tab:augmentation}

\small
\renewcommand{\arraystretch}{0.925}
\setlength{\tabcolsep}{5pt}

\begin{tabularx}{\columnwidth}{@{}p{0.38\columnwidth} c r r r@{}}
\toprule
\textbf{Training Data} & \textbf{\#Session}
& \multicolumn{1}{c}{\texttt{MRR}}
& \multicolumn{1}{c}{\texttt{nDCG@1}}
& \multicolumn{1}{c}{\texttt{nDCG@3}} \\
\midrule
Baseline & 500 & 59.12 & 39.87 & 51.76 \\
\midrule
Baseline + academic traits + research interests & 1000 & \textbf{64.37} & \textbf{48.45} & 56.92 \\
Baseline + academic traits & 1000 & 64.01 & 47.58 & \textbf{57.13} \\
Baseline + research interests & 1000 & 62.47 & 44.72 & 52.54 \\
\bottomrule
\end{tabularx}
\end{table}

\subsubsection{Results.}
Table~\ref{tab:augmentation} shows the results on SUSS Session using different augmentation methods with Agent4DL. All methods with augmented data improve over the baseline of training on the original SUSS Session training set. This shows that Agent4DL can effectively address data scarcity in sparse data scenarios by augmenting the original behavioral data to improve performance.

\section{Insights and Exploration}
Given Agent4DL's promising simulation capabilities, we pose another profound research question: \textbf{(RQ2)} Can Agent4DL provide some insights on unresolved problems in the digital library domain?

In this section, we discuss two insights drawn from our simulation results: replicating the information overload phenomenon and exploring the impact of open access on resource usage. We acknowledge the difficulty of RQ2, and confine our discussion here to potential paths for future exploration.

\subsection{Information Overload Effect}

\paragraph{\textbf{Motivation.}} Information overload is a pervasive challenge in digital libraries~\cite{bawden2020information}. This issue emerges when information retrieval systems present users with an overwhelming amount of resources, making it difficult for them to find relevant information efficiently. Our primary goal is to assess Agent4DL's ability to replicate the information overload phenomenon.

\paragraph{\textbf{Setting.}} To simulate the information overload effect, we designed an experiment using a digital library search engine API. We conducted four simulation rounds, each representing an increase in the volume of searchable resources. We conducted four simulation rounds, each employing different strategies to increase the volume and diversity of retrieved resources:

\begin{enumerate}[label=(\arabic*),labelindent=0pt,leftmargin=2em,labelsep=0.4em] 
    \item \textit{Query expansion:} We progressively expanded the initial query with related terms and synonyms in each round.
    \item \textit{Relaxing filters:} We gradually relaxed search filters (e.g., publication date range, document types) with each round.
    \item \textit{Increasing results per page:} We requested an increasing number of results per page (within API limits) in subsequent rounds.
    \item \textit{Combining multiple topics:} In later rounds, we combined queries from different but related topics to increase result diversity.
\end{enumerate}

In each round, the API returns results based on these modified parameters. We evaluate the information overload effect based on simulated user engagement at the individual level. Two metrics are employed: $\bar{T}_{\text{per-resource}}$, which represents the average time spent per accessed resource, and $\bar{N}_{\text{accessed}}$, indicating the average number of resources accessed in each simulation round.

\begin{figure}[ht]
    \centering
    \includegraphics[width=\linewidth]{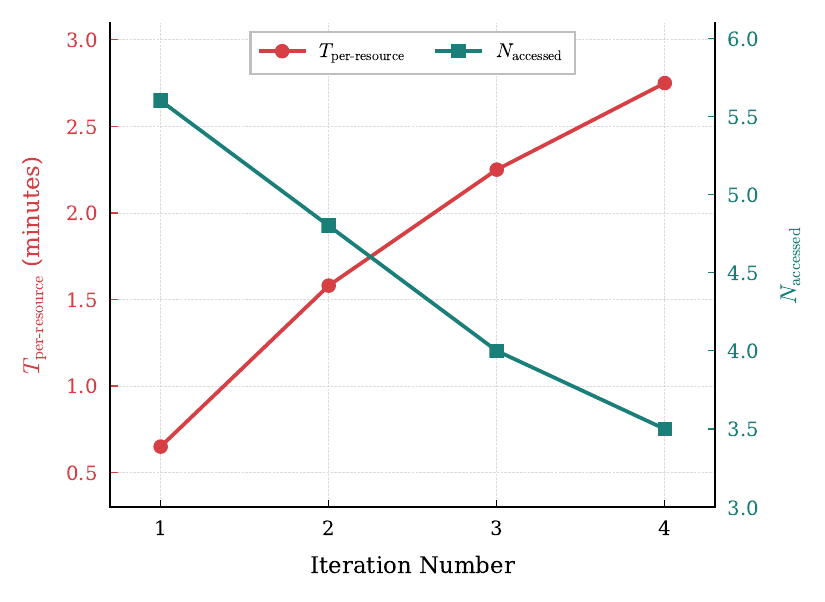}
    \caption{Information Overload Effect}
    \label{fig:user_patterns}
\end{figure}

\paragraph{ \textbf{Results.}} Figure~\ref{fig:user_patterns} reveals that as the number of iterations increases, user engagement with individual resources tends to decrease. Specifically, the time spent per resource, represented by $\bar{T}_{\text{per-resource}}$, decreases, while the number of resources accessed, denoted by $\bar{N}_{\text{accessed}}$, increases initially but then plateaus or slightly decreases. This result hint towards Agent4DL’s capability to reflect the information overload effect, an issue commonly observed in real-world digital libraries.

\subsection{Impact of Open Access on Resource Usage}

\paragraph{\textbf{Motivation.}} Understanding the impact of open access on resource usage is crucial for digital libraries to make informed decisions about their collection development and access policies. A digital library simulator can aid researchers with data collection and in addressing latent confounding issues. In light of this, a question arises: can Agent4DL be instrumental in uncovering the relationships between open access and resource usage in digital libraries?

\paragraph{\textbf{Setting.}} To understand the factors influencing resource usage in digital libraries, for each resource, we collect data on four principal variables in addition to its usage frequency simulated by agents: resource quality and relevance (sourced from the resource profile), open access status, and the number of times the resource is cited (sourced from the simulator). To probe the potential relationships within this simulated data, we employ the DirectLiNGAM algorithm [51]. This algorithm discovers a causal graph, i.e., a weighted directed acyclic graph (DAG) in a linear system.

\begin{figure}[ht]
    \centering
    \includegraphics[width=\linewidth]{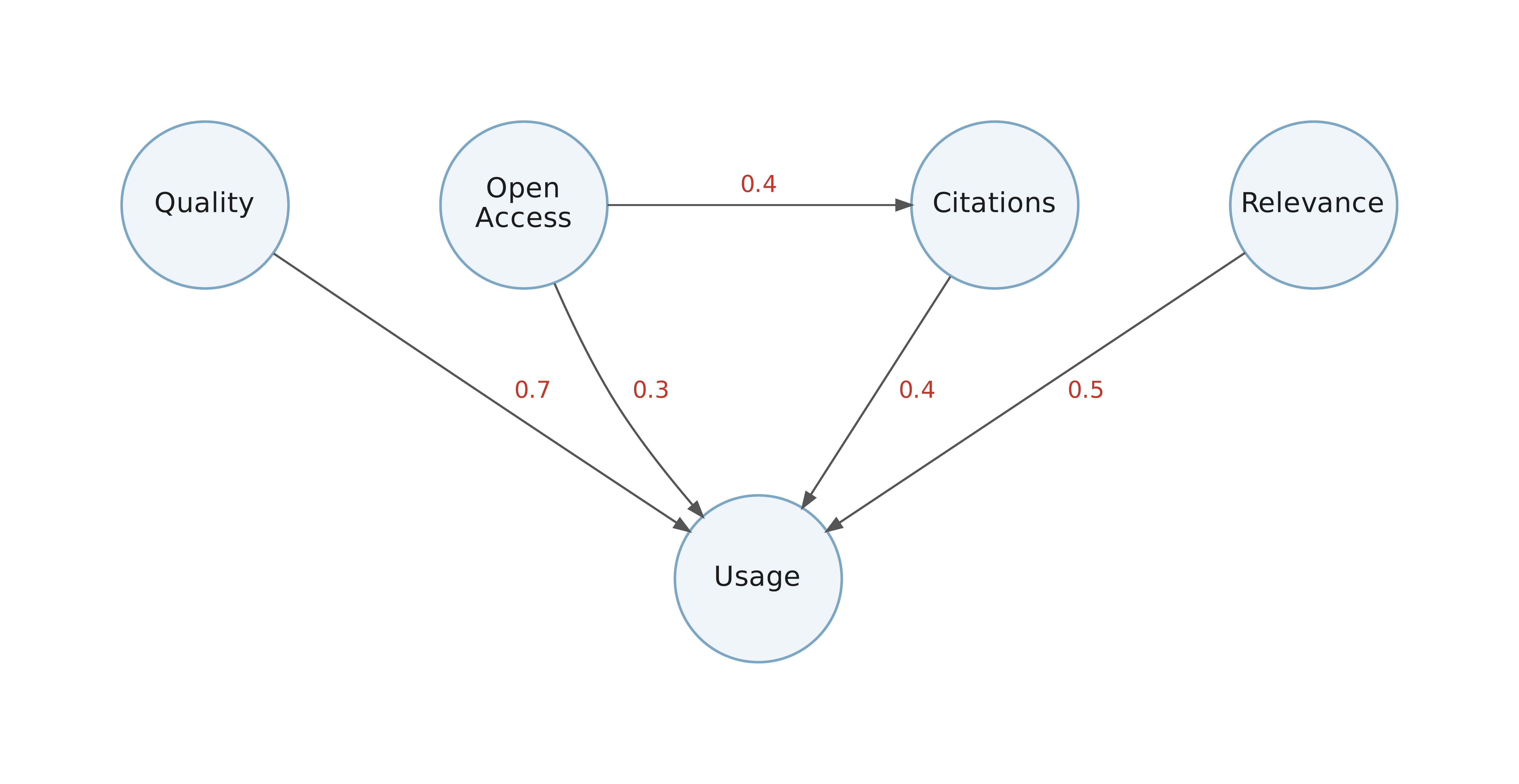}
    \caption{Impact of Open Access on Resource Usage}
    \label{fig:causal}
\end{figure}

\paragraph{ \textbf{Results.}} Based on the learned causal graph in Figure~\ref{fig:causal}, we can observe that resource quality and relevance are the primary factors influencing resource usage~\cite{pearl2016causal}. While resource quality is the most significant contributor, the relevance of resources to the user's research interests also plays a role, aligning with real-world researcher behavior. Additionally, a feedback loop amplifying the open access advantage is observed, where open access resources gain increased visibility, leading to more views and citations by agents. When these highly-used items are incorporated into new training datasets, information retrieval systems tend to rank them higher in subsequent iterations, resulting in the open access citation advantage\cite{HeWC0ZC022}. These insights suggest Agent4DL's potential for simulating complex digital library phenomena and providing valuable information for library managers and information retrieval system designers.




\section{Conclusion}
In this study, we introduced Agent4DL, a new user simulator for simulating large-scale user search behaviors in digital libraries with LLM-based agents. We designed strategies to construct user profiles that lead to human-like search behavior data. Experiments demonstrated Agent4DL's effectiveness in simulating authentic and diverse user behaviors and its potential in improving information retrieval tasks while respecting user privacy, especially in sparse datasets. We also developed Agent4DLData, a compact yet comprehensive collection of simulated search user behavior data to facilitate related research. We believe that this work can provide new perspectives to investigate the search behaviors of DL users, thereby contributing to the advancement of search technologies and user experience optimization. 


\section{Limitations and Future work}
Agent4DL presents a promising approach for simulating user behavior in digital libraries, but it has several limitations. The model relies heavily on rich metadata about academic resources, which is not uniformly available across digital libraries, limiting the ability to generate realistic queries and evaluate relevance. Additionally, the current action space of Agent4DL does not fully capture the complexity of scholarly information seeking, omitting factors like citation chaining, consultation with colleagues, or serendipitous discovery through browsing. 

We also observed instances where the language model struggled to accurately represent domain-specific knowledge or recent developments in rapidly evolving fields, leading to the generation of anachronistic queries or failure to properly assess the relevance of cutting-edge research. 

To address these issues, future work should explore techniques for augmenting sparse metadata, developing privacy-preserving methods to incorporate actual usage patterns, expanding the model to incorporate additional information pathways, and fine-tuning the language model specifically for academic information seeking, incorporating up-to-date citation databases and field-specific ontologies to improve the accuracy and currency of simulated behaviors.

\bibliographystyle{ACM-Reference-Format}
\bibliography{paper-lit}

\end{document}